\newcommand{\water}{H$_2$O}
\newcommand{\solum}{\hbox{L$_{\odot}$}}
\newcommand{\solmass}{\hbox{M$_{\odot}$}}

\newcommand{\kms}{km s$^{-1}$}

\documentclass{iaus}
\usepackage{graphicx}

\title[AGN and Megamasers] 
{AGN and Megamasers
}

\author[Andrea Tarchi]   
{Andrea Tarchi}

\affiliation{INAF - Osservatorio Astronomico di Cagliari, Capoterra (CA), Italy \\ email: {\tt atarchi@oa-cagliari.inaf.it}}

\pubyear{2012}
\volume{287}  
\pagerange{119--126}
\jname{Cosmic Masers: From OH to H$_{0}$}
\editors{R.S. Booth, E.M.L. Humphreys \& W.H.T. Vlemmings, eds.}
\begin{document}

\maketitle

\begin{abstract}
Luminous extragalctic masers are traditionally referred to as 
 the `megamasers'. Those produced by water molecules are associated with 
 accretion disks, radio jets, or outflows in the nuclear regions of 
 active galactic nuclei (AGN). The majority of OH maser sources are instead 
 driven by intense star formation in ultra-luminous infrared 
 galaxies, although in a few cases the OH maser emission traces rotating 
 (toroidal or disk) structures around the nuclear engines of AGN. Thus, 
 detailed maser studies provide a fundamental contribution to our 
 knowledge of the main nuclear components of AGN, constitute unique 
 tools to measure geometric distances of host galaxies, and have a great 
 impact on probing the, so far, paradigmatic Unified Model of AGN.

\end{abstract}

\firstsection 
\section{Introduction}\label{intro}

The widely accepted Unified Model of active galactic nuclei (AGN; e.g., \cite{antonucci93}, \cite{urry95}) implies, in their very centres, the presence of a supermassive black hole surrounded by a parsec-scale accretion disk. The emission from the disk is particularly intense at ultraviolet (UV) and soft X-ray wavelengths. The accretion disk is then surrounded by a torus (or a thick disk) of atomic and molecular gas, with a size of 1-100 pc, that obscures the optical and UV emission along certain directions. Therefore, the object appears as either a type 1 or type 2 AGN depending on the line of sight. In type 1 AGN, the observer views the accretion disk and black hole through the hole in the torus, while in type 2 AGN the direct view of these nuclear components is obscured by the torus. The amount of radio loudness in each object (thus, if it is classified as a radio-quiet or radio-loud AGN) and its membership to an individual radio class of AGN (e.g., QSO, FR\,I, BL Lac, etc...) are instead more ascribable to the host galaxy type and/or to intrinsic properties of the nuclear components of the AGN (spin, mass, and accretion rate of the black hole, the relativistic jet power and orientation, etc...; see, e.g., \cite{urry95}).

Studies of the central regions of AGN are complicated by the extremely small scales and complex structures of the nuclear components. In addition, particularly in type 2 AGN, the inner regions are often obscured at optical and UV wavelengths. Observations at infrared (IR, the band where most of the nuclear radiation absorbed by the torus is re-emitted), X-ray, and radio frequencies can, however, access these obscured regions. In particular, at radio wavelengths, water and OH maser studies are a unique tool for investigating the structure and kinematics of the gas close to and around the nuclear engines of AGN.

\section{Masing molecules in extragalactic objects}\label{molecules}

The most common molecules found to produce maser emission in extragalactic environments are hydroxyl and water. The former molecule is detected at radio wavelengths in four hyperfine transitions of its ground rotational level with rest frequencies, in the radio band, of 1665, 1667, 1612, 1720 MHz. The emission from the first two transitions is the one typically observed in galaxies. The OH maser emission traces a relatively warm (100 $<$ $T_{kin}$ $<$ 300 K) and dense (10$^{4}$ $<$ $n$(H$_{2}$) $<$ 10$^{6}$ cm$^{-3}$) gas.
The water maser main line is instead produced by the transition between the rotational levels $6_{16}$ and $5_{23}$ at a rest radio frequency of 22.2 GHz. Extragalactic water maser emission at 183 GHz was also detected in two galaxies, NGC\,3079 (\cite{humphreys05}) and Arp\,220 (\cite{cernicharo06}), that, however, will not be discussed in the following. The water maser emission traces much warmer ($T_{\rm {kin}}$ $>$ 300\,K) and denser (10$^{7}$ $<$ $n$(H$_{2}$) $<$ 10$^{11}$ cm$^{-3}$) gas than OH and the emitting spots show extreme compactness and brightness temperatures.

While the association of water megamaser sources with AGN activity has been confidently assessed for a large number of galaxies, the association between OH megamasers and AGN processes is, so far, less understood. Hence, the remainder of this review will mostly focus on H$_{2}$O megamasers, although a brief description of OH megamasers and, in particular, on their association with the H$_{2}$O ones, will be reported in Sect.~\ref{brother}. More issues related to extragalactic OH masers will also be treated elsewhere in this volume.

\begin{figure}[b]
\begin{center}
 \includegraphics[width=8cm]{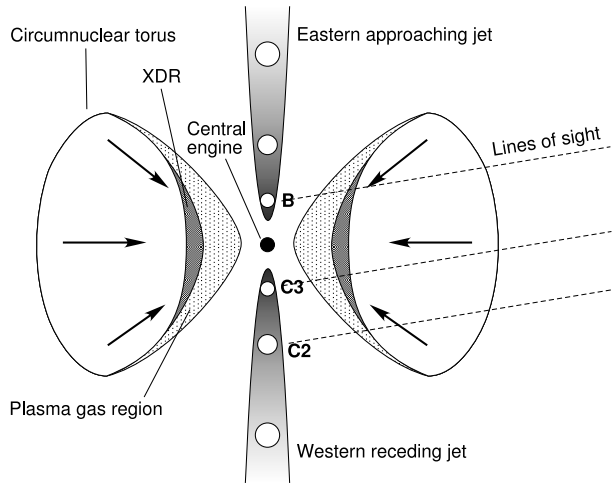} 
 \caption{A cartoon illustrating the possible environment in the circumnuclear torus and jets in NGC\,1052. An X-ray dissociation region (XDR) is formed on the inner layer of the torus and amplifies background continuum emission from the jet knots. On B and C3, we can see both the H$_{2}$O maser emission and FFA absorption. Only FFA appears, instead, on C2. The infalling of the gas inside the torus toward the central engine motivates the redshifted velocity of the water (\cite{sawada08}).}
 \label{fig1}
\end{center}
\end{figure}

\subsection{Disk masers}\label{disks}
When water masers are associated with accretion disks in AGN they often show a spectrum with a characteristic triple-peak pattern, with three distinct groups of features, one (the systemic lines) clustered around the systemic velocity of the galaxy and the other two (the high-velocity lines) almost symmetrically displaced from the first group toward the blue and the red sides by hundreds of km/s. Very Long baseline Interferometry (VLBI) and single-dish monitoring studies in the radio have allowed to use the water maser spots to map nuclear accretion disks and provide, for NGC\,4258, a calibration of the cosmic distance scale (\cite{miyoshi95}; \cite{herrnstein99}). For many years, NGC~4258 has been the only galaxy where such detailed studies were possible. However, nowadays, the number of galaxies found to host water masers associated with accretion disks and with favourable characteristics (e.g., proximity, disk inclination, etc...) is rapidly rising thanks, in particular, to the Megamaser Cosmology Project (MCP). The description of this project and its main results achieved, so far, on the use of disk-masers to derive distances of the host galaxies, hence with a strong impact on improving the accuracy with which H$_{0}$ is known, are described by Henkel and Impellizzeri et~al. (these proceedings) and will not be repeated here.

Another very relevant use of disk-masers is that, by modelling and analysing the keplerian rotation curve of the accretion disks, as derived from the aforementioned maser studies, mass estimates of the nuclear engine can be obtained. Recently, \cite{kuo11} has estimated 6 new BH masses using this method, obtaining values from 0.75 to 6.5 $\times$ 10$^{7}$ \solmass. For 3 of these galaxies, they  also derived central densities large enough to rule out clusters of stars or stellar remnants as the central objects, thus reinforcing the 'standard' supermassive black hole scenario for AGN. Furthermore, using the BH mass estimates of \cite{kuo11}, in addition to a few other BH masses estimated before using disk-masers, \cite{greene10} improved the low mass end of the M$_{BH}$ -- $\sigma^{*}$ relation, so far, almost uniquely derived for elliptical galaxies with larger BH masses. This way, they found a larger scatter in this relation than previously obtained, possibly hinting at a non-universal nature of the relation as instead indicated before by the law derived from elliptical galaxies only.

\subsection{Jet masers}\label{jets}
 Water maser emission can also be associated with radio jets, produced by either the interaction between the radio jet and an encroaching molecular cloud or by the amplification of the radio continuum from the jet from excited water molecules in a foreground cloud. Two examples of jet-maser sources (but more cases are reported in literature) can be briefly described to illustrate the two mechanisms: Mrk\,348 and NGC\,1052. In both cases, the maser spectrum is composed of a single broad (a few 100 km/s) line redshifted w.r.t. the systemic velocity of the host galaxy (for the single-dish detection spectra of Mrk\,348 and NGC\,1052, see \cite{falcke00} and \cite{braatz94}, respectively), and the maser emission was confidently found to be located along the radio continuum of the jet , displaced from the position of the putative nucleus (for Mrk\,348, see \cite{peck03}; for NGC\,1052, see \cite{claussen98}). 

Mrk\,348 was extensively studied by \cite[Peck et~al. (2003)]{peck03} using interferometric and single-dish multi-epoch observations. Their analysis led to a model for the origin of the emission related to a post-shocked region at the interface between the energetic jet material and the molecular gas in the cloud where the jet is boring through. With the aid of the reverberation mapping technique, \cite[Peck et~al. (2003)]{peck03} were also able to derive relevant physical quantities of the jet material, like its velocity and density.

A somewhat different scenario is instead that brought about for NGC\,1052 by \cite[Sawada-Satoh et~al. (2008)]{sawada08}. According to their picture, the maser clouds are most likely located foreground to the jet in a circumnuclear torus (or disk), thus amplifying the continuum seed emission from the jet knots (Fig.~\ref{fig1}). The maser gas is indeed found where the free-free opacity from foreground thermal plasma absorbing the jet synchrotron emission is large. Being the material in the torus/disk also the putative source of accretion onto the nucleus, its contraction toward the central engine accounts for the redshifted velocity of all the maser features. 

\subsection{Outflow masers}\label{outflows}
A third class of AGN-associated water masers is named 'outflow-masers'. Presently, however, there is only one case that has been thoroughly investigated, the nearby Seyfert\,2 galaxy Circinus. VLBI maps of the water emission from this galaxy have shown that maser spots trace two different dynamic components, a warped edge-on accretion disk and a wide angle nuclear outflow up to $\sim$ 1 pc from the central engine. Indeed, that in the Circinus galaxy represents the first direct evidence of dusty, high-density, molecular material in a nuclear outflow, at such small scales, thus allowing detailed study of the velocity and geometry of nuclear winds (\cite{greenhill03}). At larger scales,  more recently, the distribution of dust in the nuclear region of Circinus has been investigated by \cite{tristram07} using interferometric observations with the MID-infrared Interferometric Instrument (MIDI) at the VLTI.  They found that the dust is distributed in two components, a dense and warm disk component with a radius of 0.2 pc (where the disk-masers found by \cite{greenhill03} are located) and a less dense and slightly cooler geometrically-thick torus-like components up to pc scales. This dusty torus confirms the presence of such structures in AGN, as expected from the Unified Model, and is seemingly the agent that collimates the AGN outflow traced by the water maser spots. However, evidence is also reported by \cite{tristram07} of a clumpy or filamentary dust distribution in the torus that, if confirmed for AGN tori in general, may have an impact on our understanding/classification of the different types of AGN. 

\section{Water masers in AGN: detection rates and host galaxies}\label{rates}

So far, more than 3000 galaxies have been searched for water maser emission and  detections have been obtained in about 150 of them\footnote{These values are compiled using information taken from the Megamaser Cosmology Project (MCP) and Hubble Constant Maser Experiment (HoME), `https://safe.nrao.edu/wiki/bin/view/Main/MegamaserCosmologyProject' and `https://www.cfa.harvard.edu/~lincoln/demo/HoME/index.html', respectively.}, the majority being radio-quiet AGN classified as Seyfert\,2 (Sy\,2) or Low-ionization nuclear emission-line regions (LINERs), in the local Universe (z $<$ 0.05) 

The overall detection rate for AGN is only of a few \%. This detection rate rises considerably (up to 20-25\%) when considering only the nearest Sy\,2 and LINER galaxies (e.g. those $<$ 5000 km/s; Braatz et al. in prep.).

A question that naturally arises is: why some AGN, also among the same class, host H$_{2}$O maser sources while some other don't? One way to try answering this question is to investigate possible peculiarities of the masing galaxies (w.r.t. the non-masing ones).

Indeed, AGN hosting water maser emission tend to show a high column density ($N_{\rm H} > 10^{23}\,{\rm cm}^{-2}$) or are even Compton-thick ($N_{\rm H} > 10^{24}\,{\rm cm}^{-2}$; e.g. \cite{zhang06}, \cite{greenhill08}). Furthermore, a rough correlation has been found between maser isotropic luminosity and unabsorbed X-ray luminosity (\cite{kondratko06}). Although promising, these studies have been affected by a lack or poor quality of X-ray data for a large percentage ($\sim$ 50\%) of the known maser galaxies. Positively, a survey of all known H$_{2}$O maser sources in AGN using the Swift satellite is ongoing (\cite{castangia11}, Castangia et~al. in prep.) that may help clarifying, on a firm statistical basis, the interplay between X-ray and maser emission.

A number of studies have been also recently led with the goal of finding correlations between the occurrence of maser emission and host galaxies characteristics at several wavelengths. The main results of these studies are: the confirmation of the fact that H$_{2}$O masers reside primarily in X-ray absorbed sources (\cite{zhang10}, \cite{ramolla11}, \cite{zhu11}), the larger radio luminosity found in H$_{2}$O maser hosts w.r.t. that in non-maser galaxies (\cite{zhang12}), and the indication of the extinction-corrected [OIII]$\lambda$5007 flux as a good sample selection criterion to maximize the number of detections in maser searches (\cite{zhu11}; see also Zaw et~al., these proceedings). 

As already mentioned before, the highest occurrence of maser emission happens in type\,2 Seyferts. Surely, identifying water maser emission associated with Sy\,2 galaxies is not particularly surprising, given that the Unified Model for AGN requires an obscuring structure, which is probably an edge-on disk or torus, along the line of sight towards the nucleus of a type 2 AGN. This structure can provide a molecular reservoir and the amplification paths necessary for maser action. The relation between a type 1 Seyfert and the maser phenomenon is instead less obvious. According to the same paradigm, in type 1 objects, accretion disks and/or tori, if present at all, should be orientated face-on, making them less likely to produce detectable maser emission. Consistently, out of the 150 water maser sources detected so far, maser emission has been detected in only one 'pure' Sy\,1, NGC\,2782, apparently in strong agreement with the Unified Model. However, a recent result obtained by \cite[Tarchi et~al. 2011b]{tarchi11b} seems to complicate the aforementioned conclusion. 

Indeed, in the last years, a (sub)class of Seyfert\,1 (Sy\,1) galaxies, named Narrow-Line Seyfert\,1 (NLS1s), has attracted particular attention among astronomers. NLS1s have the broad emission-line optical spectra of type 1 Seyfert galaxies, but with the narrowest Balmer lines from the broad line region and the strongest Fe II emission (e.g., \cite{osterbrock89}, \cite{veron01}). Extreme properties are also observed in X-rays (see \cite{komossa08} for a review). By including all past surveys (Braatz et~al. in prep) and the outcome of two new surveys, \cite[Tarchi et~al. (2011b)]{tarchi11b} has compiled a list of all NLS1s (71) searched, so far, in the water maser line. Five successful detections are reported, yielding a detection rate of $\sim$ 7\%, which is comparable with the global rates of AGN surveys. While this result is surprising, the maser detection rate in NLS1s becomes even more impressive when we consider a volume-limited sample that somewhat minimizes the limitation in sensitivity of the survey(s). When considering only NLS1 galaxies at recessional velocities less than 10000 \kms, the detection rate goes up dramatically to $\sim$ 21 \% (5/24). This value approaches the highest detection rates ever obtained for similarly volume-limited samples, in any class of AGN. Possible explanations for such a high water maser detection rate may reside in the NLS1s’ peculiar properties: accretion rates close to their Eddington limit, putative small black hole masses, strong radiation-pressure outflows, and viewing angles intermediate between type 1 and type 2 Seyferts. 

\subsection{Maser in Elliptical and/or Radio-loud galaxies}\label{loud}
So far, (almost) no water masers have been found in elliptical and/or radio-loud galaxies despite a number of surveys has been performed on different classes of AGN. 

\begin{figure}[b]
\begin{center}
 \includegraphics[width=8cm]{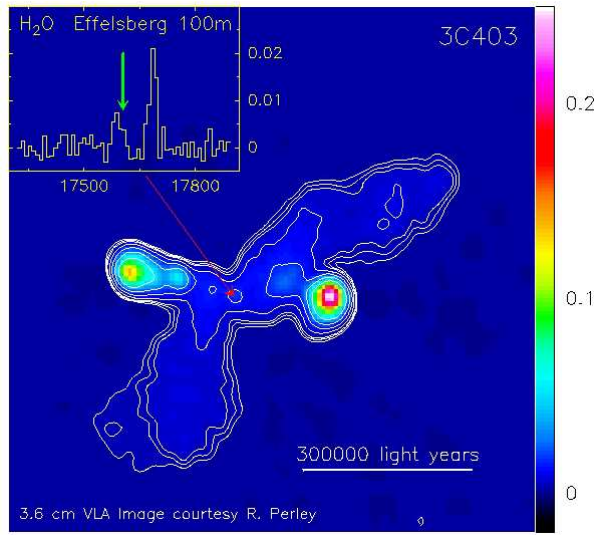} 
 \caption{NRAO Very Large Array image of the radio galaxy 3C\,403 at a wavelength of 3.6 cm. The intensity range of the colors (in Jansky, Jy, units) is indicated at the right hand side. The red arrow points at the galaxy's nucleus. The 22-GHz water maser spectrum shown in the upper left hand inset was taken with the Effelsberg 100m telescope. The green arrow points at the systemic radial velocity of the whole galaxy (Credits: National Radio Astronomy Observatory/Rick Perley (NRAO/AUI/NSF); the water maser spectrum is taken from \cite{tarchi03}).}
 \label{fig2}
\end{center}
\end{figure}

Among the AGN targeted in these surveys are:

\begin{itemize}
\item 50 Fanaroff-Riley I (FR\,I) with $z$ $<$ 0.15 (\cite{henkel98})
\item 273 Type 2 QSOs with 0.3 $<$ $z$ $<$ 0.83 (\cite{bennert09})
\item 79 radio galaxies, mostly Fanaroff-Riley II (FR\,II) with $z$ $<$ 0.17  (GBT project \#AGBT02C$\_$030, unpublished)
\item 5 Grav. Lensed Quasars with 2.3 $<$ $z$ $<$ 2.9 (\cite{mckean11})
\item 17 Type 1 QSOs with $z$ $<$ 0.06 (\cite{koenig12})\\
\end{itemize}

No new water maser detections was reported.\\

There are, however, few exceptions:

$-$ The radio galaxy NGC\,1052 at z=0.005 (\cite{claussen98}; \cite{sawada08}). The maser source in this galaxy, discovered by \cite[Braatz et~al. (1994)]{braatz94}, has been already described in Sect.~\ref{jets}.

$-$ The FR\,II galaxy 3C\,403 at z=0.06. This galaxy displays a peculiar X-shaped radio morphology (Fig.~\ref{fig2}). The maser emission showed two main broad lines experiencing extreme flux variability within about a year period. In view of linewidths and the lack of satellite lines, an interpretation in terms of an association of the maser emission with the radio jets is the most plausible one (\cite{tarchi03}; \cite{tarchi07b})

$-$ The type 2 QSO SDSS\,J0804+3607 at z=0.66. The maser is composed of a single feature, redshifted w.r.t. to the systemic velocity of the galaxy, with isotropic luminosity of 23000 \solum. No other maser features are reported. So far, the nature of the maser emission has not been assessed (\cite{barvainis05}; \cite{bennert09})

$-$ The type 1 quasar MG\,J0414+0534 at z=2.64. The most distant and luminous ($\sim$ 30000 \solum) maser source and, so far, the only one found in a gravitational lensed system (\cite{impellizzeri08}; \cite{castangia11}). For a description of this source, see also Castangia et~al. (these proceedings).

$-$ The NLS1  IGR\,16385-2057  at z=0.03. Indeed, while the nature of the galaxy hosting the maser emission has still to be confidently determined, IGR16385-2057 is optically classified as an elliptical and displays a core+lobes radio morphology, resembling that of classical radio galaxies (\cite{tarchi11b}; Castangia et al. in prep.).\\

Despite the small number of detections does not allow us to derive any definite conclusion, we can note that the detection of masers in E/radio-loud objects seem not to be very much AGN-type dependent. Indeed, so far, there have been detections both in type 1 and type 2 objects. Generally, masers in E/radio-loud objects are more seemingly associated with radio-jets or outflows, although the disk-maser scenario cannot be a priori ruled out. Obviously, VLBI observations are necessary to better assess the origin of the emission. However, these measurements have been successfully performed only for the nearest source, NGC\,1052, due to the intrinsic weakness or flaring-down of the maser lines in the other cases.

Among the possible interpretations for the paucity of water maser detections in elliptical and/or radio-loud galaxies, we can report: i) the possible lack of molecular material (e.g. \cite{henkel98}); ii) instability due to tidal disruption of molecular maser clouds in circumnuclear disks orbiting around black holes that are particularly massive, as those in powerful radio galaxies (e.g., \cite{tarchi07b}); iii) a steep or non-evolving water maser luminosity function that would not provide us with maser sources luminous enough to be detectable at the large distances at which the elliptical/radio-loud galaxies are located (this option seems, however, to be ruled out by the recent work of \cite{mckean11}, where an indication for a slow - at least - evolution of the maser luminosity function with $z$ is reported); iv) strong variability of the maser features, as seen, for example in 3C\,403 (on the other hand, very high flux stability has been reported by \cite{castangia11} for the maser in MG\,J0414+0534); v) the insufficient sensitivity of the surveys led so far (in this framework, the enhanced sensitivity of some of the present and upcoming radio facilities will help to quantify the true relevance of this item).

\section{H$_{2}$O and OH megamasers: brotherhood or mutual exclusion?}\label{brother}

\subsection{OH megamasers in AGN}\label{oh}
OH megamasers (so far, about 100 sources are known) are almost uniquely associated with LIRGs ($L_{\rm IR} > 10^{11}$ \solum) and ULIRGs ($L_{\rm IR} > 10^{11}$ \solum), are IR radiatively pumped, and a relation between OH and FIR luminosities exists, $L_{\rm OH} \propto   L_{\rm FIR}^{1.2}$ (e.g., \cite{baan89}; \cite{darling02}).

The OH emission is typically the contribution of two components. One component is diffuse and was explained as low-gain, unsaturated amplification of background radio continuum by foreground clouds (e.g., \cite{baan85}). The second component is compact and produced by high-gain saturated emission (e.g., \cite{lonsdale02}). A model explaining both emissions by a single gas phase has been more recently described in Parra et al. (2005; see also the following text).

\begin{figure}[b]
\begin{center}
 \includegraphics[width=8cm]{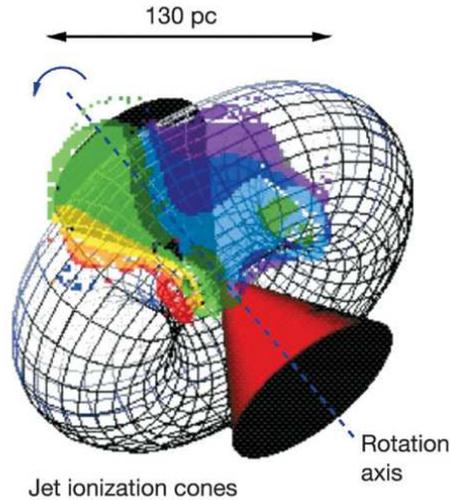} 
 \caption{The inferred model of the nuclear torus in Mrk~231 is displayed as a wire diagram with symmetric ionization cones. This model takes into account all large--scale characteristics of the nuclear radio emission and the OH emission. The molecular material moves from top--right to bottom--left, i.e., northwest to southeast (\cite{kloeckner03}).}
 \label{fig3}
\end{center}
\end{figure}

The region of maser emission has an extent of up to 100 pc and it is typically found in high concentration of molecular gas where intense/extreme starbursts are ongoing. It has been speculated that the association with starbursts may indicate that the OH megamaser occurrence in a galaxy is a short-lived phenomenon (e.g., Lo 2005, and references therein)

Despite, as just mentioned, OH megamasers seem to have a star-formation origin, two cases have been reported where a connection between the OH megamaser and AGN activity is present,  Mrk\,231
 and IIIZw\,35. In the former source, the OH maser emission (mapped with the VLBI) traces a rotating, dusty, molecular torus (or thick disk) located between 30 and 100 pc from the central engine (\cite{kloeckner03}; Fig.~\ref{fig3}). Similarly, in IIIZw\,35, the OH emission is produced by pc-size OH maser clouds (overlapping) in the tangent points of a nearly edge-on torus (\cite{pihlstroem01}). For this source, both the diffuse and compact emission component have been successfully explained by a single phase of unsaturated clumpy gas in a ring structure amplifying background continuum (\cite{parra05}).

Another OH megamaser source that has recently attracted a lot of attention is that in Arp\,299, a merger system produced by the close interaction of two galaxies, IC\,694 and NGC\,3690. OH megamaser emission was detected from the nuclear region of IC\,694 (\cite{baan85}). The OH emission was found coincident in position with the radio continuum peak and confined in a 100 pc rotating structure (\cite{polatidis01}). In the same galaxy, water maser emission was also detected (\cite{tarchi07a}), with a position slightly offset w.r.t. that of the OH maser, and associated with an expanding slab of material, seemingly a nuclear outflow (\cite{tarchi11a}). While VLBI observations of the maser emission from both molecules are planned, interestingly, \cite{perez10} has found strong evidence for the presence of the long-sought putative AGN in the system, associated with the nucleus of IC\,694.

\subsection{On the OH and H$_{2}$O maser relation}\label{relation}
By compiling all sources where searches for maser emission from both \water\ and OH has been, so far, reported in the literature, it has been found that, out of the resulting 51 galaxies, 33 galaxies show only water maser emission, 13 galaxies show only OH maser emission, and 5 sources confidently show maser emission from both molecules (\cite{tarchi11a}; their Fig.~11). In particular, the two well-known starburst galaxies NGC\,253 and M\,82 show weak maser emission from \water\ and OH, the two water megamaser galaxies NGC\,3079 and NGC\,1068 also host a weak OH maser, and Arp\,299 shows megamaser emission from both molecules. Thus, only Arp\,299 shows the contemporary presence, in IC\,694 (see previous paragraph), of luminous OH and H$_{2}$O maser emission (\cite{surcis09}; \cite{tarchi11a}). 

What is then possibly causing this apparent lack of contemporary detections in the same object of megamaser emission from both molecules (duration of the maser phenomena, merger stage in some systems, other)? To answer this question a more systematic approach (parallel surveys to detect \water\  maser emission in OH maser galaxies, and viceversa) are necessary. Furthermore, a more detailed study of the, so far, unique case, IC\,694, is auspicable since it offers the possibility to investigate different physical conditions and dynamic structures in the same AGN.


\section{Concluding remarks}\label{summary}

The results obtained, so far, strengthen the uniqueness of the contribution provided by megamaser studies to our understanding of the nuclear regions of AGN. H$_{2}$O megamasers are preferentially found in nearby (radio-quiet) type 2 Seyferts with some exceptions (e.g., NLS1s), somewhat consistently, with the AGN Unified Model. H$_{2}$O megamasers are, instead, rarely found in elliptical and radio-loud galaxies with only few special cases. In this case, given the low statistics, a connection with the Unified Model is more difficult. In addition, OH megamasers are preferentially found in ULIRGs associated with extreme star formation rather than AGN activity. Nevertheless, in few cases, OH emission traces rotating disk/torus structures as those invoked by the Unified Model for AGN. A mutual exclusion between OH and  H$_{2}$O megamasers seems also to be present, with Arp\,299 as the only exception.

While a large number of OH and water maser sources is under detailed investigation, the overall understanding of the maser phenomenon in the framework of the standard Unified Model for AGN is still far from being assessed, mostly due to the lack of maser detections and/or detailed-enough observations in the radio-loud domain of AGN.

Furthermore, a number of open issues still needs to be clarified. Among these, the most relevant are the presently-unknown beaming of the maser radiation and the degree of clumpiness of the material in the disk/torus. These two elements may have a relevant influence on the true luminosity estimates and excitation requirements of the maser sources, and on the classification of the AGN type hosting the maser emission.

A final mention needs to be done on the existence of alternative or revised versions of the standard Unified Model for AGN (e.g., \cite{nenkova08}; \cite{elitzur12}). When trying to derive a comprehensive picture of the AGN `zoo', observational and theoretical results should necessarily be compared with these models as well. Once again, in this framework, maser studies offer among the best tools to test both the standard and alternative scenarios.

\begin{acknowledgements}
The author would like to thank Paola Castangia for critically reading the manuscript and Jim Braatz for providing useful information prior to publication.

\vspace{0.15cm}
\centering{\it {This review is dedicated to the memory of Albert Greve, a mentor and a friend.}}
\end{acknowledgements}


\begin{thebibliography}{}

\bibitem[{{Antonucci} 1993}]{antonucci93}
{Antonucci}, R. 1993, 
\textit{ARAA}, 31, 473

\bibitem[{{Baan} 1985}]{baan85}
{Baan}, W.~A. 1985, 
\textit{Nature}, 315, 26

\bibitem[{{Baan} 1989}]{baan89}
{Baan}, W.~A. 1989, 
\textit{ApJ}, 338, 804

\bibitem[{{Barvainis \& Antonucci} 2005}]{barvainis05}	
{Barvainis}, R., \& {Antonucci}, R. 2005,
\textit{ApJ}, 628, L89

\bibitem[{{Bennert} {et~al.} 2009}]{bennert09}
{Bennert}, N., {Barvainis}, R., {Henkel}, C., \& {Antonucci}, R. 2009, 
\textit{ApJ}, 695, 276


\bibitem[{{Braatz} {et~al.} 1994}]{braatz94}
{Braatz}, J.~A., {Wilson}, A.~S., \& {Henkel}, C. 1994, 
\textit{ApJ}, 437, L99

\bibitem[{{Castangia} {et~al.} 2010}]{castangia10}
{Castangia}, P., {Tilak}, A., {Kadler}, M., {Henkel}, C., {Greenhill}, L., \& {Tueller}, J. 2010,
\textit{X-ray Astronomy 2009} Proc. AIPC, Vol. 1248, p.\ 347

\bibitem[{{Castangia} {et~al.} 2011}]{castangia11}
{Castangia}, P., {Impellizzeri}, C.~M.~V.., {McKean}, J.~P., {Henkel}, C., {Brunthaler}, A., {Roy}, A.~L., {Wucknitz}, O., {Ott}, J., \& {Momjian}, E. 2011
\textit{A\&A}, 529, 150

\bibitem[{{Cernicharo, Pardo \& Wei{\ss}} 2006}]{cernicharo06}
{Cernicharo}, J., {Pardo}, J.~R., \& {Wei{\ss}}, A., 2006
\textit{ApJ}, 646, L49

\bibitem[{{Claussen} {et~al.} 1998}]{claussen98}
{Claussen}, M.~J., {Diamond}, P.~J., {Braatz}, J.~A., {Wilson}, A.~S., \& {Henkel}, C. 1998
\textit{ApJ}, 500, L129

\bibitem[{{Darling \& Giovanelli} 2002}]{darling02}
{Darling}, J., \& {Giovanelli}, R. 2002,
\textit{ApJ}, 572, 810


\bibitem[{{Elitzur} 2012}]{elitzur12}
{Elitzur}, M. 2012,
\textit{ApJ}, 747, L33

\bibitem[{{Falcke} {et~al.} 2000}]{falcke00}
{Falcke}, H., {Henkel}, C., {Peck}, A.~B., {Hagiwara}, Y., {Prieto}, M.~A., \& {Gallimore}, J.~F. 2000
\textit{A\&A}, 358, 17

\bibitem[{{Greene} {et~al.} (2010)}]{greene10}
{Greene}, J.~E., {Peng}, C.~Y., {Kim}, M., {Kuo}, C.~-Y., {et~al.} 2010,
\textit{ApJ}, 721, 26

\bibitem[{{Greenhill} {et~al.} 2003}]{greenhill03}
{Greenhill}, L.~J., {Booth}, R.~S., {Ellingsen}, S.~P., {et~al.} 2003,
\textit{ApJ}, 590, 162

\bibitem[{{Greenhill} {et~al.} 2008}]{greenhill08}
{Greenhill}, L.~J., {Tilak}, A., \& {Madejski}, G. 2008, 
\textit{ApJ}, 686, L13

\bibitem[{{Henkel} {et~al.} 1998}]{henkel98}
{Henkel}, C., {Wang}, Y.~P., {Falcke}, H., {Wilson}, A.~S., \& {Braatz}, J.~A. 1998,
\textit{A\&A}, 335, 463

\bibitem[{{Herrnstein} {et~al.} 1999}]{herrnstein99}
{Herrnstein}, J.~R., {Moran}, J.~M., {Greenhill}, L.~J., {Diamond}, P.~J., {Inoue}, M., {Nakai}, N., {Miyoshi}, M., {Henkel}, C., \& {Riess}, A. 1999,
\textit{Nature}, 400, 539

\bibitem[{{Humphreys} {et~al.} 2005}]{humphreys05}
{Humphreys}, E.~M.~L., {Greenhill}, L.~J., {Reid}, M.~J., {Beuther}, H., {Moran}, J.~M., {Gurwell}, M., {Wilner}, D.~J., \& {Kondratko}, P.~T. 2005,
\textit{ApJ}, 634, 133

\bibitem[{{Impellizzeri} {et~al.} 2008}]{impellizzeri08}	
{Impellizzeri}, C.~M.~V., {McKean}, J.~P., {Castangia}, P., {Roy}, A.~L., {Henkel}, C., {Brunthaler}, A., \& {Wucknitz}, O. 2008,
\textit{Nature}, 456, 927

\bibitem[{{Kl\"{o}ckner, Baan, \& Garrett} 2003}]{kloeckner03}
{Kl\"{o}ckner}, H.-R., {Baan}, W.~A., \& {Garrett}, M.~A. 2003,
\textit{Nature}, 421, 821

\bibitem[{{K\"{o}nig} {et~al.} 2012}]{koenig12}
{K\"{o}nig}, S., {Eckart}, A., {Henkel}, C., \& {Garc\'{i}a-Mar\'{i}n}, M. 2012
\textit{MNRAS}, 420, 2263

\bibitem[{{Komossa} 2008}]{komossa08}
{Komossa}, S. 2008,
\textit{Rev. Mexicana AyA} (Conference Series), vol.~32, 86

\bibitem[{{Kondratko, Greenhill \& Moran} 2006}]{kondratko06}
{Kondratko}, P.~T., {Greenhill}, L.~J.; \& {Moran}, J.~M. 2006,
\textit{ApJ}, 652, 136

\bibitem[{{Kuo} {et~al.} (2011)}]{kuo11}
{Kuo}, C.~Y., {Braatz}, J.~A., {Condon}, J.~J., {et~al.} 2011, 
\textit{ApJ}, 727, 20

\bibitem[{{Lo} 2005}]{lo05}
{Lo}, K.~Y. 2005, 
\textit{ARAA}, 43, 625

\bibitem[{{Lonsdale} 2002}]{lonsdale02}
{Lonsdale}, C.~J. 2002, 
\textit{Cosmic Masers: From Proto-Stars to Black Holes}. Proc. IAU Symposium No. 206 (San Francisco: ASP), p.\ 413

\bibitem[{{McKean} {et~al.} 2011}]{mckean11}
{McKean}, J.~P., {Impellizzeri}, C.~M.~V., {Roy}, A.~L., {Castangia}, P., {Samuel}, F., {Brunthaler}, A., {Henkel}, C., \& {Wucknitz}, O. 2011,
\textit{MNRAS}, 410, 2506

\bibitem[{{Miyoshi} {et~al.} 1995}]{miyoshi95}
{Miyoshi}, M., {Moran}, J., {Herrnstein}, J., {Greenhill}, L., {Nakai}, N., {Diamond}, P., \& {Inoue}, M. 1995, 
\textit{Nature}, 373, 127

\bibitem[{{Nenkova} {et~al.} 2008}]{nenkova08}
Nenkova, M., Sirocky, M.~M., Nikutta, R., Ivezić, Z., \& Elitzur, M., 2006,
\textit{ApJ}, 685, 160

\bibitem[{{Osterbrock} 1989}]{osterbrock89}
{Osterbrock}, D.~E. 1989,
\textit{Astrophysics of gaseous nebulae and active galactic nuclei}, ed.\ University Science Books


\bibitem[{{Parra} {et~al.} 2005}]{parra05}
{Parra}, R., {Conway}, J.~E., {Elitzur}, M., \& {Pihlstr\"{o}m}, Y.~M. 2005,
\textit{A\&A}, 443, 383

\bibitem[{{Peck} {et~al.} 2003}]{peck03}
{Peck}, A.~B., {Henkel}, C., {Ulvestad}, J.~S., {Brunthaler}, A., {Falcke}, H., {Elitzur}, M., {Menten}, K.~M., \& {Gallimore}, J.~F. 2003, 
\textit{ApJ}, 590, 149

\bibitem[{{P\'{e}rez-Torres} {et~al.} (2010)}]{perez10}
P\'{e}rez-Torres, M.~A., Alberdi, A., Romero-Canizales, C., \& Bondi, M. 2010,
\textit{A\&A}, 519, L5

\bibitem[{{Pihlstr\"{o}m} {et~al.} 2001}]{pihlstroem01}
Pihlstr\"{o}m, Y.~M., Conway, J.~E., Booth, R.~S., Diamond, P.~J., \& Polatidis, A.~G. 2001,
\textit{A\&A}, 377, 413

\bibitem[{{Polatidis \& Aalto} 2001}]{polatidis01}
Polatidis, A.~G., Aalto, S. 2001,
\textit{Galaxies and their Constituents at the Highest Angular Resolutions}. Proc. IAU Symposium No. 205 (San Francisco: ASP), p.\ 1

\bibitem[{{Ramolla} {et~al.} 2011}]{ramolla11}
{Ramolla}, M., {Haas}, M., {Bennert}, V.~N., \& {Chini}, R. 2011, 
\textit{A\&A}, 530, 147

\bibitem[{{Sawada-Satoh} {et~al.} 2008}]{sawada08}
{Sawada-Satoh}, S., {Kameno}, S., {Nakamura}, K., {Namikawa}, D., {Shibata}, K.~M., \& {Inoue}, M. 2008, 
\textit{ApJ}, 680, 191

\bibitem[{{Surcis} {et~al.} 2009}]{surcis09}
{Surcis}, G., {Tarchi}, A., {Henkel}, C., {Ott}, J., {Lovell}, J., \& {Castangia}, P. 2009,
\textit{A\&A}, 502, 529

\bibitem[{{Tarchi} {et~al.} 2003}]{tarchi03}
{Tarchi}, A., {Henkel}, C., {Chiaberge}, M., \& Menten, K.~M. 2003, 
\textit{A\&A}, 407, L33

\bibitem[{{Tarchi} {et~al.} 2007a}]{tarchi07a}
{Tarchi}, A., {Castangia}, P., {Henkel}, C., \& {Menten}, K.~M. 2007a,
\textit{New Astron. Revs}, 51, 67

\bibitem[{{Tarchi} {et~al.} 2007b}]{tarchi07b}
{Tarchi}, A., {Brunthaler}, A., {Henkel}, C., {Menten}, K.~M., {Braatz}, J., \& {Wei\ss}, A. 2007b, 
\textit{A\&A}, 475, 497

\bibitem[{{Tarchi} {et~al.} 2011a}]{tarchi11a}
{Tarchi}, A., {Castangia}, P., {Henkel}, C., {Surcis}, G., \& {Menten}, K.~M. 2011a, 
\textit{A\&A}, 525, 91

\bibitem[{{Tarchi} {et~al.} 2011b}]{tarchi11b}
{Tarchi}, A., {Castangia}, P., {Columbano}, A., {Panessa}, F., \& {Braatz}, J.~A. 2011b,
\textit{A\&A}, 532, 125

\bibitem[{{Tristram} {et~al.} (2007)}]{tristram07}
{Tristram}, K.~R.~W., {Meisenheimer}, K., {Jaffe}, W., {et~al.} 2007,
\textit{A\&A}, 474, 837

\bibitem[{{Urry} \& {Padovani} 1995}]{urry95}
{Urry}, C.~M. \& {Padovani}, P. 1995, 
\textit{PASP}, 107, 803

\bibitem[{{V{\'e}ron-Cetty} {et~al.} 2001}]{veron01}
{V{\'e}ron-Cetty}, M., {V{\'e}ron}, P., \& {Gon{\c c}alves}, A.~C. 2001, 
\textit{A\&A}, 372, 730

\bibitem[{{Zhang} {et~al.} 2006}]{zhang06}
{Zhang}, J.~S., {Henkel}, C., {Kadler}, M., {Greenhill}, L.~J., {Nagar}, N., {Wilson}, A.~S., \& {Braatz}, J.~A. 2006, 
\textit{A\&A}, 450, 933

\bibitem[{{Zhang} {et~al.} 2010}]{zhang10}
{Zhang}, J.~S., {Henkel}, C., {Guo}, Q., {Wang}, H.~G., \& {Fan}, J.~H. 2010,
\textit{ApJ}, 708, 1528

\bibitem[{{Zhang} {et~al.} 2012}]{zhang12}
{Zhang}, J.~S., {Henkel}, C., {Guo}, Q., \& {Wang}, J. 2012,
\textit{A\&A}, 538, 152

\bibitem[{{Zhu} {et~al.} 2011}]{zhu11}
{Zhu}, G., {Zaw}, I., {Blanton}, M.~R., \& {Greenhill}, L.~J. 2011,
\textit{ApJ}, 742, 73

\end{thebibliography}
\end{document}